\documentclass[onecolumn,trackchanges]{aastex631}
\usepackage{amsmath}
\shorttitle{Synchrotron polarization of anisotropic electron distribution in GRB prompt emission}
\shortauthors{Cheng et al.}
\graphicspath{{./}{figures/}}
\begin{document}

\title{Synchrotron polarization of anisotropic electron distribution in GRB prompt emission}

\correspondingauthor{Jirong Mao}
\email{jirongmao@mail.ynao.ac.cn}

\author[0009-0004-3324-8421]{Kang-Fa Cheng}
\affiliation{Guangxi Key Laboratory for Relativistic Astrophysics, School of Physical Science and Technology, Guangxi University \\
Nanning 530004, People’s Republic of China}
\affiliation{School of Physics and Electronic Information, Guangxi Minzu University \\
Nanning 530006, People’s Republic of China}

\author{Kai-Xian Luo}
\affiliation{School of Physics and Electronic Information, Guangxi Minzu University \\
Nanning 530006, People’s Republic of China}

\author{Xiao-Hong Zhao}
\affiliation{Yunnan Observatories, Chinese Academy of Sciences \\ 
Kunming, 650011, People’s Republic of China}
\affiliation{Center for Astronomical Mega-Science, Chinese Academy of Science \\
20A Datun Road, Chaoyang District, Beijing 100012, People’s Republic of China}

\author{Jirong Mao}
\affiliation{Yunnan Observatories, Chinese Academy of Sciences \\ 
Kunming, 650011, People’s Republic of China}
\affiliation{Center for Astronomical Mega-Science, Chinese Academy of Science \\
20A Datun Road, Chaoyang District, Beijing 100012, People’s Republic of China}

\author{Hong-Bang Liu}
\affiliation{Guangxi Key Laboratory for Relativistic Astrophysics, School of Physical Science and Technology, Guangxi University \\
Nanning 530004, People’s Republic of China}

\author{Yu-Hang Mo}
\affiliation{School of Physics and Electronic Information, Guangxi Minzu University \\
Nanning 530006, People’s Republic of China}

\author{Jin-Rong Huang}
\affiliation{School of Physics and Electronic Information, Guangxi Minzu University \\
Nanning 530006, People’s Republic of China}

\author{Rong-Li Weng}
\affiliation{School of Physics and Electronic Information, Guangxi Minzu University \\
Nanning 530006, People’s Republic of China}

\author{Wen-Jie Xie}
\affiliation{School of Physics and Electronic Information, Guangxi Minzu University \\
Nanning 530006, People’s Republic of China}

\author{Gao-Jin Yu}
\affiliation{School of Physics and Electronic Information, Guangxi Minzu University \\
Nanning 530006, People’s Republic of China}




\begin{abstract}
In gamma-ray bursts (GRBs), the electron pitch angle ($\alpha$) is usually assumed to be isotropically distributed. However, recent numerical simulations indicate that only the high-energy electrons (with Lorentz factors $\gamma>\gamma_{iso}$) are distributed isotropically, whereas the low-energy electrons (with $\gamma<\gamma_{iso}$) follow an energy-dependent anisotropic distribution during magnetic reconnection. The mean value of $\sin^2 \alpha$ approximately follows the relation $\langle \sin^2 \alpha \rangle \propto \gamma^{m}$ for $\gamma<\gamma_{iso}$. In principle, polarization measurements may help us constrain the pitch-angle distribution of electrons in GRBs, since different pitch-angle distributions produce distinct synchrotron polarization signatures. The polarization of GRBs produced by isotropically distributed electrons has been extensively studied. In this paper, we investigate synchrotron polarization produced by anisotropically distributed electrons within a globally toroidal magnetic field in GRB prompt emission. Our results show that the synchrotron PDs in the $\gamma$-ray and X-ray bands produced by anisotropically distributed electrons are systematically lower than those produced by isotropically distributed electrons, while the PD in the optical band could be either lower or higher than that of isotropically distributed electrons, depending primarily on the value of the energy slope $m$. In addition, we compared our numerical results with observational data, and the comparison suggests that an anisotropic distribution of electrons may offer a potential explanation for the PD and spectral data of some GRBs.
\end{abstract}



\section{Introduction} \label{sec:intro}

Gamma-ray bursts (GRBs) are the most energetic explosive phenomena in the universe, characterized by intense and highly variable emission across the electromagnetic spectrum. The prompt emission phase, typically lasting milliseconds to hundreds of seconds, is believed to originate from relativistic jets launched by central engines—either massive stellar collapse or compact object mergers \citep{1989Natur.340..126Eichler,1993ApJ...405..273Woosley}.
Despite decades of observational and theoretical efforts, the detailed radiation mechanisms and particle acceleration processes responsible for GRB prompt emission remain debated.

Synchrotron radiation from relativistic electrons accelerated in magnetic fields (MFs) is one of the most promising mechanisms for GRB prompt emission. In standard synchrotron models, electrons are typically assumed to have an isotropic pitch-angle (the angle between the direction of electron velocity and the direction of the MF) distribution. Under this assumption, the synchrotron spectrum in GRBs has been extensively studied \citep{1998ApJ...497L..17S, 2004RvMP...76.1143P, 2006RPPh...69.2259M,  Uhm+Zhang+2014, 2014ApJ...780...12Z, 2015PhR...561....1K}, and its polarization characteristics in the prompt phase have also been widely investigated \citep{Toma+etal+2009, Gill+etal+2020, Gill+Granot+2024, 2024MNRAS.52712178G, Lan+etal+2019, Lan+etal+2021(2), 2025ApJ...994..194W, 2024ApJ...975..277C, 2025ApJ...985..112C, 2025ApJ...986...32Z}. However, the observed GRB spectra often exhibit low-energy spectral indices (\(F_\nu \propto \nu^{0}\)) harder than the typical synchrotron prediction of \(F_\nu \propto \nu^{-1/2}\) under fast cooling regime \citep{1993ApJ...413..281B, 2000ApJS..126...19P, 2006ApJS..166..298K}. This is known as the "synchrotron fast cooling problem" \citep{2000MNRAS.313L...1G}.

Recent particle-in-cell simulations of relativistic magnetic reconnection have revealed that particle acceleration in reconnection layers can produce electron distributions that deviate significantly from the isotropic assumption \citep{2019ApJ...886..122C, 2021PhRvL.127y5102C, 2022ApJ...936L..27C, 2020ApJ...895L..40C, 2023ApJ...959..137C, 2024ApJ...972....9C}. \cite{2019ApJ...886..122C} showed that the low-energy electrons are accelerated along the guiding MF with small pitch angles, and the pitch angle distribution becomes isotropic at higher energies in the simulations of turbulence magnetic reconnection. Furthermore, \cite{2023ApJ...959..137C} demonstrated that reconnection-driven particle acceleration in pair plasmas imprints an energy-dependent pitch-angle ($\alpha$) anisotropy, characterized by broken power laws in both the electron energy spectrum and the mean value of \(\sin^2\alpha\) as a function of electron Lorentz factor (LF). Specifically, they found that \(\langle \sin^2\alpha \rangle\) follows a piecewise power-law behavior: \(\langle \sin^2\alpha \rangle \propto \gamma^{m_<}\) for low-energy electrons below a characteristic LF \(\gamma_{\min, \alpha}\), and \(\langle \sin^2\alpha \rangle \propto \gamma^{m_>}\) for intermediate energies up to \(\gamma_{\mathrm{iso}}\), beyond which isotropy (\(\langle \sin^2 \alpha \rangle = 2/3\)) is restored. The power-law indices \(m_<\) and \(m_>\), as well as the characteristic energies \(\gamma_{\min, \alpha}\) and \(\gamma_{\mathrm{iso}}\), depend sensitively on the ratio of guide field to reconnecting field \((B_g/B_0)\) and the lepton magnetization \((\sigma_0)\). Extending this work to ion-electron plasmas, \cite{2024ApJ...972....9C} showed that similar energy-dependent pitch-angle anisotropy emerges for both species, with the energy partition between ions and electrons strongly regulated by \(B_g/B_0\). These findings establish that anisotropic electron pitch-angle distributions are a generic outcome of magnetic reconnection in magnetically dominated plasmas, with potential implications for GRB prompt emission.

The presence of energy-dependent pitch-angle anisotropy can significantly modify both the spectral shape and polarization properties of synchrotron radiation. \cite{2018ApJ...864L..16Y} adopt a Gaussian distribution in the $\sin \alpha$ space to calculate the synchrotron spectra. Compared to the typical synchrotron line of death of \(F_\nu \propto \nu^{1/3}\) derived from the isotropic distribution electrons, they obtained a new synchrotron line of death of  \(F_\nu \propto \nu^{2/3}\). \cite{2022ApJ...933...18G} considered a Gaussian distribution in the $\cos \alpha$ space to calculate the synchrotron spectra, and reproduced the typical GRB spectrum under specific parameters. \cite{2023ApJ...959..137C} derived that for a power-law electron distribution \(N(\gamma) \propto \gamma^{-p}\) combined with pitch-angle anisotropy \(\langle \sin^2 \alpha \rangle \propto \gamma^{m}\). The radiation power spectrum becomes \(F_\nu \propto \nu^{(2-2p+m)/(4+m)}\) compared to the standard expression \(F_\nu \propto \nu^{(1-p)/(2)}\) , while the intrinsic polarization degree (PD) becomes \(\Pi_0 = (p+1)/(p+7/3 + m/3)\) compared to the standard isotropic result \(\Pi_0 = (p+1)/(p+7/3)\) \citep{Rybicki+Lightman+1979}. The electron energy slope of \(m\) determines whether the spectral indices and PDs are enhanced (\(m<0\)) or suppressed (\(m>0\)) relative to the isotropic case. However, these intrinsic polarization estimates do not account for the geometry effect and MF configuration. If we apply these results to GRBs, we should consider the influence of the jet geometry and MF configuration on the observed polarization.
\citep{Granot+2003, Gill+Granot+2021, Cheng+etal+2024, 2025ApJ...988..202M, 2025ApJ...994..194W}.

In this paper, we investigate the synchrotron polarization properties of GRB prompt emission produced by anisotropically distributed electrons, motivated by the numerical findings of \cite{2023ApJ...959..137C} and \cite{2024ApJ...972....9C}. Our goal is to distinguish the electron pitch-angle distribution by combining the polarization of GRBs with their spectral properties. Adopting a globally toroidal MF configuration within a uniform top-hat jet model, and incorporating the energy-dependent pitch-angle anisotropy, we calculate the PDs across $\gamma$-ray, X-ray, and optical bands, and examine their dependence on the anisotropy parameters and the viewing angles. This paper is organized as follows. Section \ref{sec:dis-ele} describes the anisotropic electron distribution. Section \ref{sec:cal-pol} shows the calculations of synchrotron polarization with anisotropic electron distribution in GRB prompt emission, including the energy-resolved polarization, the distribution of polarization with the energy slopes, the distribution of polarization with viewing angles, and the comparision with the measured PD and spectral data. The summary and discussions are presented in Section \ref{sec: conclusion}.

\section{Anisotropic electron distribution in GRB prompt emission} \label{sec:dis-ele}

Although often assumed to be isotropic, the electron pitch-angle distribution may become anisotropic due to magnetic reconnection, as suggested by recent simulations \citep{2020ApJ...895L..40C, 2023ApJ...959..137C, 2024ApJ...972....9C}. This is manifested as an approximately isotropic distribution of high-energy electrons, whereas the pitch-angle distribution of low-energy electrons is energy-dependent.
Electrons with LFs above $\gamma_{iso}$ are distributed isotropically, whereas those below $\gamma_{iso}$ follow an anisotropic distribution. $\langle \sin^{2} \alpha \rangle$ (the mean value of $\sin^{2} \alpha$) approximately follows the relation as  \citep{2023ApJ...959..137C, 2024ApJ...972....9C}

\begin{equation} \label{ele-alpha}
\langle \sin^{2} \alpha \rangle \approx
\begin{cases}

\Lambda (\gamma/\gamma_{min, \alpha})^{m_{<}}   & (\rm{for \quad \gamma_{min} \leq \gamma \leq \gamma_{min, \alpha}}) \\
\Lambda (\gamma/\gamma_{min, \alpha})^{m_{>}}   & (\rm{for \quad  \gamma_{min, \alpha} \leq \gamma \leq \gamma_{iso}}) \\
2/3  & (\rm{for \quad \gamma_{iso} \leq \gamma \leq \gamma_{max}})

\end{cases}
\end{equation} 

where $\alpha$ is the electron pitch angle. $\gamma$ is the LF of the electron. $\Lambda \sim min \langle \sin^2 \alpha \rangle$. $m_<$ and $m_>$ are the negative-slope and positive-slope energy dependence of the pitch-angle anisotropy, respectively.  $\gamma_{min, \alpha}$ is the LF at which the pitch-angle anisotropy is strongest. $\gamma_{iso}$ is the minimum LF of the isotropic distribution of electrons. $\gamma_{min}$ and $\gamma_{max}$ are the minimum and maximum  LFs of the electrons, respectively. We take $\gamma_{\rm{min}}=3$ and $\gamma_{max}=10^7$ in this paper. Due to the small pitch angle of the anisotropically distributed electrons, we can adopt the approximate relationship $\langle \sin \alpha \rangle \sim \sqrt{\langle \sin^{2} \alpha \rangle}$ for $\gamma <\gamma_{iso}$, then we obtain

\begin{equation} \label{ele-pa}
\langle \sin \alpha \rangle \approx
\begin{cases}

S_0 (\gamma/\gamma_{min, \alpha})^{m_{1}}   & (\rm{for \quad \gamma_{min} \leq \gamma \leq \gamma_{min, \alpha}}) \\
S_0 (\gamma/\gamma_{min, \alpha})^{m_{2}}   & (\rm{for \quad  \gamma_{min, \alpha} \leq \gamma \leq \gamma_{iso}}) \\
\pi/4  & (\rm{for \quad \gamma_{iso} \leq \gamma \leq \gamma_{max}})

\end{cases}
\end{equation} 

where $S_0 \sim min \langle \sin \alpha \rangle$, we take $S_0 =0.01$ in this paper. $m_1$ ($m_1=m_{<}/2$) and $m_2$ ($m_2=m_{>}/2$) are the negative-slope and positive-slope energy dependence of the pitch-angle anisotropy, respectively.  According to the principle of continuity, the following relation should hold

\begin{equation} 
S_0 (\gamma_{iso}/\gamma_{min \alpha})^{m_2}=\pi/4
\end{equation} 

thus we obtain 
\begin{equation} \label{g_iso}
\gamma_{iso}=\gamma_{min, \alpha} (\frac{\pi}{4 S_0})^{\frac{1}{m_2}} 
\end{equation} 

\begin{figure}[ht!]
\plotone{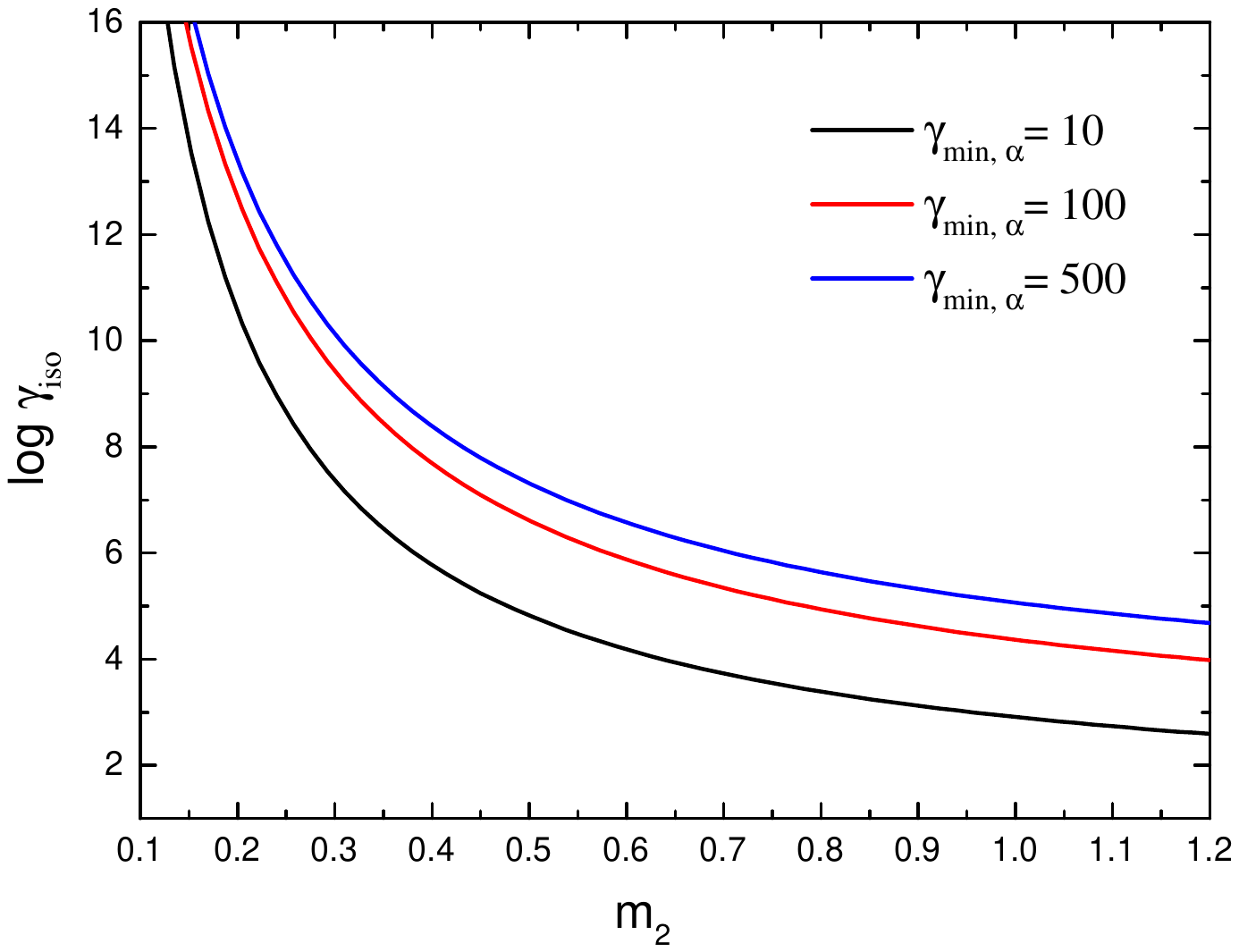}
\caption{The curves of $log \gamma_{iso}$ versus $m_2$ with different $\gamma_{min, \alpha}$. \label{fig: m2-giso}}
\end{figure}

The relationship between $m_2$ and $\gamma_{iso}$ is shown in Figure \ref{fig: m2-giso}. This figure show that as $m_2$ increases, $\gamma_{iso}$ declines rapidly, while the decrease is slower when $\gamma_{min, \alpha} $ is larger.
Following the numerical results of \cite{2023ApJ...959..137C}, the electron energy spectrum is distributed as a broken power law, which can be described as 
\begin{equation} \label{ele-dis}
N_{e} (\gamma) =
\begin{cases}
N_{0} (\gamma/\gamma_0)^{-p_1}  & (\rm{for \quad \gamma_{min}\leq \gamma \leq \gamma_{0}}) \\
N_{0} (\gamma/\gamma_0)^{-p_2}  & (\rm{for \quad \gamma_{0} \leq \gamma \leq \gamma_{max}})
\end{cases}
\end{equation} 

where $p_1$ and $p_2$ are the segmented power-law indices of the electron energy spectrum. $N_{0}$ is a normalization constant. $\gamma_0$ is the conjunctive LF.

\section{Calculations of synchrotron polarization with anisotropic electron distribution in GRB prompt emission } \label{sec:cal-pol}

The electron pitch-angle distribution provides important insights into particle acceleration mechanisms in jets and the radiation processes of GRBs.
The radiation spectra produced by anisotropically distributed electrons have been studied \citep{2018ApJ...864L..16Y, 2022ApJ...933...18G, 2023ApJ...959..137C, 2024ApJ...972....9C},  but research on the polarization resulting from such electron distributions remains limited \citep{ 2023ApJ...959..137C, 2024ApJ...972....9C}. Polarization observations of GRBs can serve as a probe to constrain the pitch-angle distribution of electrons. In this paper, we consider jet geometric effects and magnetic field structures to calculate the polarization of GRBs produced by anisotropically distributed electrons, and compare it with the polarization generated by isotropically distributed electrons. We consider a uniform top-hat jet within a globally toroidal MF model in this paper. For isotropically distributed electrons ($\gamma_{iso} \leq \gamma \leq \gamma_{max}$), the PD of an electron population with a distribution of $N_{e} (\gamma)$ can be calculated as \citep{Rybicki+Lightman+1979}
\begin{equation} \label{pi0}
\Pi_0=\frac{\int_{\gamma_{\rm{iso}}}^{\gamma_{\rm{max}}}  G(x)  N_{e} (\gamma) d\gamma}{\int_{\gamma_{\rm{iso}}}^{\gamma_{\rm{max}}}  F(x)  N_{e} (\gamma) d\gamma}
\end{equation}

where $F(x)$ and $G(x)$ are written as \citep{Rybicki+Lightman+1979}
\begin{equation}
\begin{cases}
F(x)=x\int_{x}^{\infty}K_{5/3}(\xi)d\xi \\
G(x)=xK_{2/3}(x),
\end{cases}
\end{equation}
where $K_{5/3}(\xi)$ and $K_{2/3}(x)$  are Bessel functions, $x=\nu'/\nu'_{c}$, and $\nu'_c=\frac{3q_{e}B'\sin\alpha}{4\pi m_e c}\gamma^{2}$. Note that $B'$ and $\nu'$ are the MF strength and the frequency in the jet comoving frame. If the electron spectral index is p, the analytical expression of equation \ref{pi0} can be written as $\Pi_0= \frac{p+1}{p+7/3}$ \citep{Rybicki+Lightman+1979}. Considering the jet geometric effect and MF configuration, the linear PD of isotropically distributed electrons ($\gamma_{iso} \leq \gamma \leq \gamma_{max}$) of GRB prompt emission in the toroidal MF model can be calculated as \citep{Cheng+etal+2020}
\begin{equation} \label{pi-nu}
\begin{aligned}
\Pi (\nu)&=\frac{Q(\nu)}{I(\nu)}= \int_{0}^{(1+q)^{2}y_{j}} g(y) dy \int_{-\Delta \phi(y)}^{\Delta \phi(y)} d\phi  \int_{\gamma_{\rm{iso}}}^{\gamma_{\rm{max}}} G(x)N_{e}(\gamma) \\
&\times B' \sin \alpha \cos (2\chi ) d\gamma  [ \int_{0}^{(1+q)^{2}y_{j}} g(y) dy\\
&\times \int_{-\Delta \phi(y)}^{\Delta \phi(y)} d\phi  \int_{\gamma_{\rm{iso}}}^{\gamma_{\rm{max}}} F(x) N_{e}(\gamma) B'  \sin \alpha  d\gamma]^{-1}.
\end{aligned}
\end{equation}
where $Q(\nu)$ and $I(\nu)$ are the stokes parameters. Note that $U(\nu)=0$. Some variables are defined here: $y \equiv (\Gamma \theta)^{2}$,  $y_{j} \equiv (\Gamma \theta_{j})^{2}$, and $q \equiv \theta_{v} / \theta_{j}$. where $\Gamma$ is the bulk LF of the jet, $\theta_{j}$ is the jet opening angle, and $\theta_{v}$ is the viewing angle. $\gamma_{\rm{max}}$ is the maximum LF of the electrons. For the time-averaged PD, $g(y)=(1+y)^{-2}$ \citep{Nakar+etal+2003}. By taking the electron distribution (equation \ref {ele-dis}) into equation \ref{pi-nu}, we can calculate the PD of GRB prompt emission. 
Other variables are as follows \citep{Granot+2003, Granot+konigl+2003, Granot+Taylor+2005, Toma+etal+2009}:
\begin{equation}
\sin \alpha = \left[ \left(\frac{1-y}{1+y}\right)^{2} + \frac{4y}{(1+y)^{2}} \frac{(s-\cos \phi)^{2}}{(1+s^{2}-2s \cos \phi)} \right]^{1/2},
\end{equation}
\begin{equation}
\chi = \phi + \rm{arctan} \left( \frac{(1-y)}{(1+y)} \frac{\sin \phi}{(s- \cos \phi)}\right),
\end{equation}
\begin{equation}
\Delta \phi(y) =
\begin{cases}
0,    \qquad \qquad \qquad \qquad \rm{for} \; q>1  \; \rm{and}  \;  y<(1-q)^2 y_{j}     \\
\pi,  \qquad \qquad \qquad \qquad \rm{for} \; q<1  \; \rm{and}  \;  y<(1-q)^2 y_{j}   \\
\cos^{-1} \left[\frac{(q^{2}-1)y_{j}+y}{2q \sqrt{y_{j}y}}\right]   \qquad \qquad \qquad \qquad \quad \rm{otherwise}.
\end{cases}
\end{equation}
where $s=\theta/\theta_v$. For anisotropically distributed electrons ($\gamma_{min} \leq \gamma \leq \gamma_{iso}$), the PD of an electron population with a distribution of $N_{e} (\gamma)$ is modified from equation \ref {pi0} to the expression below 
\begin{equation} \label{pi0-aniso}
\Pi_0=\frac{\int_{\gamma_{\rm{min}}}^{\gamma_{\rm{iso}}}  G(x) \langle \sin \alpha \rangle N_{e} (\gamma) d\gamma}{\int_{\gamma_{\rm{min}}}^{\gamma_{\rm{iso}}}  F(x) \langle \sin \alpha \rangle N_{e} (\gamma) d\gamma}
\end{equation}
The analytical expression of equation \ref{pi0-aniso} can be written as $\Pi_0= \frac{p+1}{p+7/3+m/3}$ \citep{2023ApJ...959..137C}, where $p$ is the electron spectral index and $m$ is the energy slope in equation \ref{ele-alpha}. As indicated by this equation, compared with the polarization arising from isotropically distributed electrons, the parameter $m$ exerts a significant influence on the PD. However, this equation only represents the intrinsic PD, observationally, the PD of GRBs is also affected by the jet geometry and the configuration of the MF. For anisotropically distributed electrons ($\gamma_{min} \leq \gamma \leq \gamma_{iso}$), the PD in GRB prompt emission can be modified from equation \ref {pi-nu} to the expression below
\begin{equation} \label{pi-nu-aniso}
\begin{aligned}
\Pi (\nu)&=\frac{Q(\nu)}{I(\nu)}= \int_{0}^{(1+q)^{2}y_{j}} g(y) dy \int_{-\Delta \phi(y)}^{\Delta \phi(y)} d\phi  \int_{\gamma_{\rm{min}}}^{\gamma_{\rm{iso}}} G(x)N_{e}(\gamma) \\
&\times B' \langle \sin \alpha \rangle \cos (2\chi ) d\gamma  [ \int_{0}^{(1+q)^{2}y_{j}} g(y) dy\\
&\times \int_{-\Delta \phi(y)}^{\Delta \phi(y)} d\phi  \int_{\gamma_{\rm{min}}}^{\gamma_{\rm{iso}}} F(x) N_{e}(\gamma) B'  \langle \sin \alpha \rangle  d\gamma]^{-1}.
\end{aligned}
\end{equation}
Taking $\langle sin \alpha \rangle$ within the range of $\gamma_{min} \leq \gamma \leq \gamma_{iso}$ from equation \ref{ele-pa} into equation \ref{pi-nu-aniso}, we can obtain the PD contributed by the anisotropically distributed electrons  in GRB prompt emission.

\subsection{Energy dependence of polarization}\label{subsec:energy-res}
\begin{figure}[ht!]
\plotone{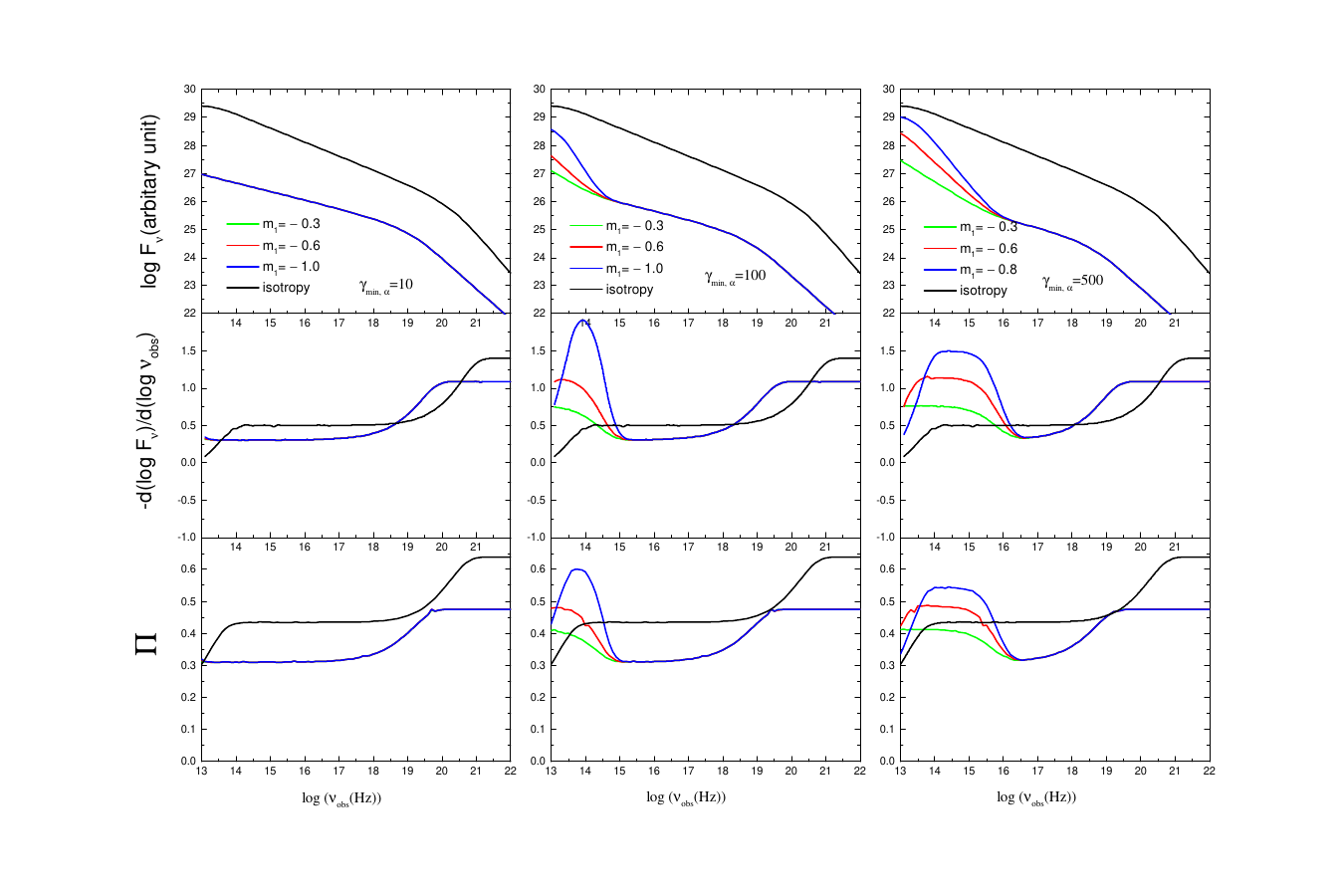}
\caption{The flux density spectra (upper panels), the negative spectral index of flux density spectra (middle row panels), and the energy-resolved PDs (bottom panels) with different energy slope of $m_1$. We fix the energy slope of $m_2=0.3$ in this figure, and take $\gamma_{min, \alpha}=10$, $100$, and $500$ in the left, middle, and right column panels, respectively. For comparison, the black solid line corresponds to the case of isotropic electron distribution.\label{fig:nu-chm1-pol}}
\end{figure}

To investigate the effect of the anisotropic distribution of electron pitch angles on the PD across various energy bands, we calculate and present the variation of time-averaged PD with energy under the anisotropic electron distribution. The flux density spectra, the negative spectral index $\alpha_0$ ($F_{\nu} \propto \nu^{-\alpha_0}$) of flux density spectra, and the energy-resolved PDs are shown with different energy slope of $m_1$  in Figure \ref{fig:nu-chm1-pol}. In this figure, we fix the energy slope of $m_2=0.3$ and vary $m_1$ to examine the effect of $m_1$ on the polarization spectrum. We take $\gamma_{min, \alpha}=10$, $100$, and $500$ in the left, middle, and right column panels, respectively. Then we find that, compared with the isotropic distribution case, both of the negative spectral index and PD spectrum display a significant bump in the optical band for $\gamma_{min, \alpha}=100$ and $500$. Specifically, the smaller the value of $m_1$, the higher the PDs in the optical band. This could be verified by conducting optical polarimetry observations of GRBs that exhibit optical flashes. While for $\gamma_{min, \alpha}=10$, since the anisotropic distribution is primarily concentrated in low-energy electrons, the influence of $m_1$ may occur in lower-energy bands than the optical band.

\begin{figure}[ht!]
\plotone{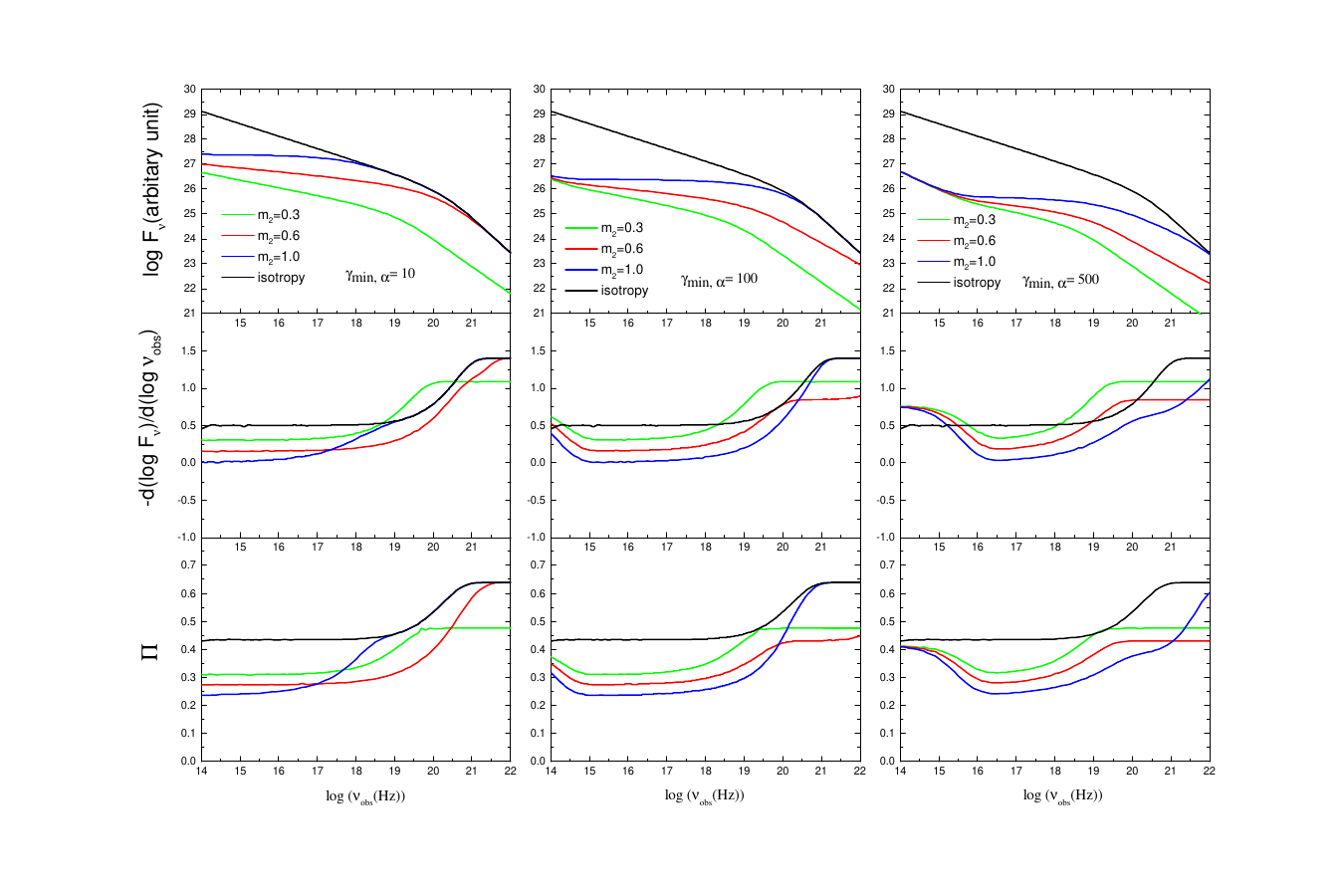}
\caption{Similar to Figure \ref{fig:nu-chm1-pol}, but with different energy slope of $m_2$. We fix the energy slope of $m_1=-0.3$ in this figure. \label{fig:nu-chm2-pol}}
\end{figure}

Similar to Figure \ref{fig:nu-chm1-pol}, for studying the influence of the energy slope of $m_2$ on the radiation spectrum and polarization of GRBs, we fixed $m_1=-0.3$ and varied the value of $m_2$ in Figure \ref{fig:nu-chm2-pol}. From the middle row of panels, it can be seen that, compared to the isotropic case, the radiation spectrum produced by anisotropically distributed electrons exhibits hardening. The low-energy negative spectral index of isotropically distributed electrons shown in the figure is $\sim 1/2$, which is consistent with the theoretical synchrotron spectrum in the fast cooling regime \citep{1998ApJ...497L..17S}. For $m_2 \approx 1$, the low-energy negative spectral index of the radiation produced by anisotropically distributed electrons approach zero, which is in agreement with the observed Band spectrum. Therefore, the anisotropic electron distribution model may provide a potential explanation for the fast cooling problem in synchrotron radiation. From the bottom panels, we find that irrespective of whether $\gamma_{min, \alpha}=10$, $100$, or $500$, the value of $m_2$ exerts a significant influence on the PDs across all wavebands of the GRBs. Compared to the case with an isotropic electron distribution, the anisotropic distribution leads to a reduction in the PDs within the $\gamma$-ray bands, X-ray, and optical bands. It should be noted that when the value of $m_2$ is large ($m_2 \gtrsim 1.0$), the PD of high-energy $\gamma$-ray radiation may approach that observed under an isotropic electron distribution. This is because a larger $m_2$ corresponds to a lower minimum LF ($\gamma_{iso}$) for the isotropically distributed electrons (see Figure \ref{fig: m2-giso}), where the high-energy $\gamma$-ray emission predominantly originates from these isotropic electrons.

\subsection{Polarization as a function of the energy slopes $m_1$ and $m_2$} \label{subsec:pol-es}
\begin{figure}[ht!] 
\plotone{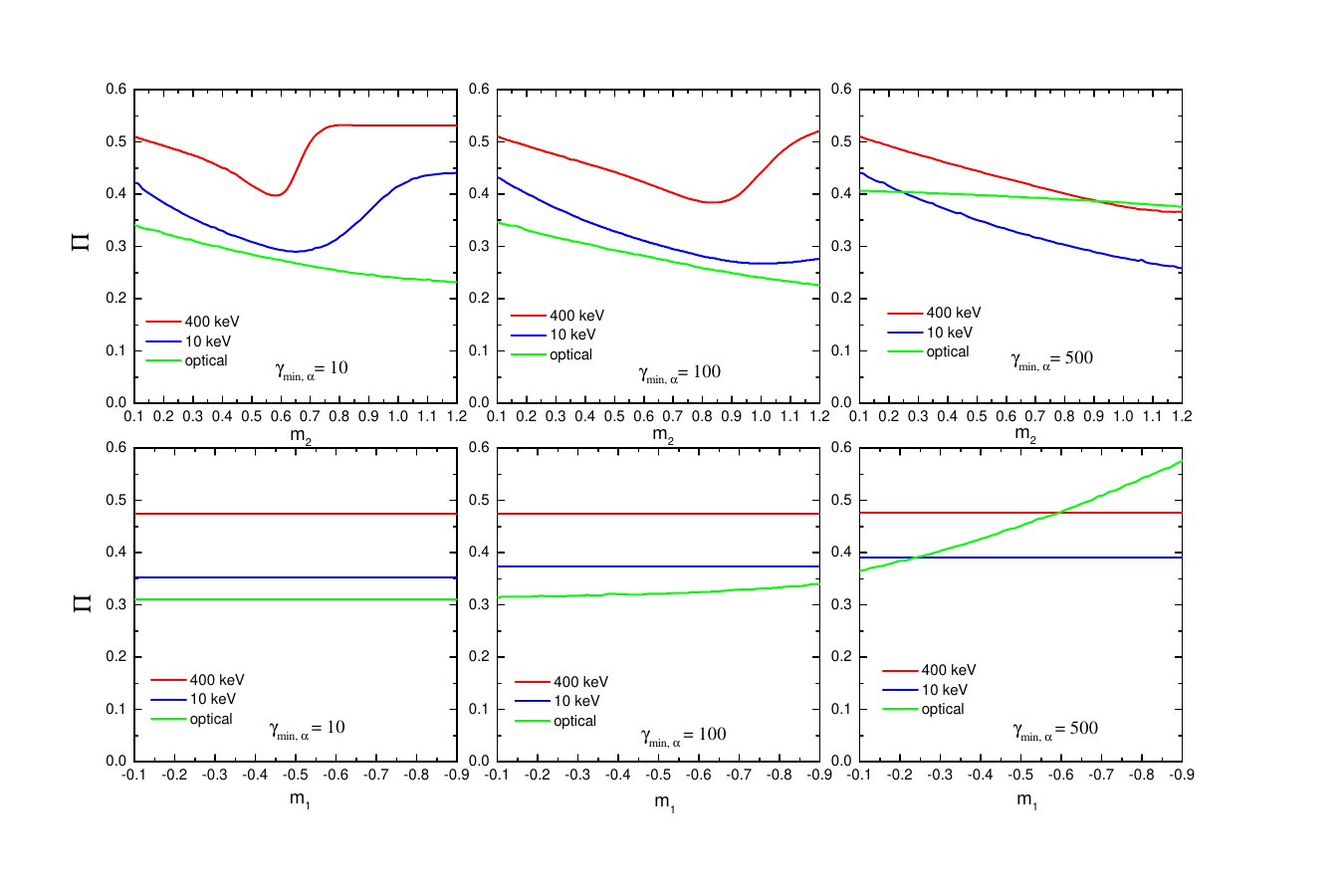}
\caption{The distribution of the PDs as a function of the energy slopes of $m_1$ (bottom panels) and $m_2$ (upper panels) in the $\gamma$-ray (400 keV), X-ray (10 keV), and optical ($6\times 10^{14}$ Hz) bands  of GRB prompt emission. We fix $m_1=-0.3$ in the upper panels and fix $m_2=0.3$ in the bottom panels. We take $\gamma_{min, \alpha}=10$, $100$, and $500$ in the left, middle, and right column panels, respectively. 
 \label{fig:res-m1-m2}}
\end{figure}

In this subsection, we will calculate the distribution of the PDs across the $\gamma$-ray (400 keV), X-ray (10 keV), and optical ($6\times 10^{14}$ Hz) bands of GRB prompt emission as a function of the energy slopes $m_1$ and $m_2$, respectively. These results are shown in Figure \ref{fig:res-m1-m2}. In the bottom panels, we find that the value of $m_1$ only affects the PD below the optical band. This is because $m_1$ corresponds to the energy slope of the low-energy electrons and its influence is confined to radiation in the low-energy regime.The influence of $m_1$ on the optical band manifests as a gradual increase in the PD as $m_1$ decreases. In the upper panels, the parameter of $m_2$ exerts a significant influence on all three energy bands of $\gamma$-ray, X-ray, and optical.The PD in the optical band gradually decreases with increasing $m_2$, whereas in the $\gamma$-ray and X-ray bands, the PD does not necessarily vary monotonically with $m_2$. For cases with relatively small $\gamma_{iso}$ values ($10$ and $100$), the PD in the $\gamma$-ray and X-ray bands initially decreases and then increases with rising $m_2$. This occurs because when $m_2$ increases, $\gamma_{iso}$ decreases rapidly (see Figure \ref{fig: m2-giso}). Once $\gamma_{iso}$ falls within the range of the electron LF where the radiation peak energy in the gamma-ray and X-ray bands, the radiation in these bands transitions from being dominated by anisotropically distributed electrons to being dominated by isotropically distributed electrons, leading to the recovery of PD. In contrast, for cases with larger $\gamma_{iso}$ values ($500$), the corresponding $\gamma_{iso}$ is higher for the same $m_2$ value (see Figure \ref{fig: m2-giso}). Consequently, the $\gamma$-ray and X-ray bands are less likely to shift from anisotropic to isotropic electron dominance, resulting in a monotonic decrease in PD with increasing $m_2$. In addition, the energy slope of $m_1$ causes an increase in PD in the optical band, whereas $m_2$ leads to a decrease. Therefore, when the influence of $m_1$ exceeds that of $m_2$, the PD in the optical band should be higher than that of isotropically distributed electrons. Conversely, it would be lower than that of isotropic distribution case.

\subsection{Polarization as a function of normalized viewing angle}\label{subsec:general}
\begin{figure}[ht!]
\plotone{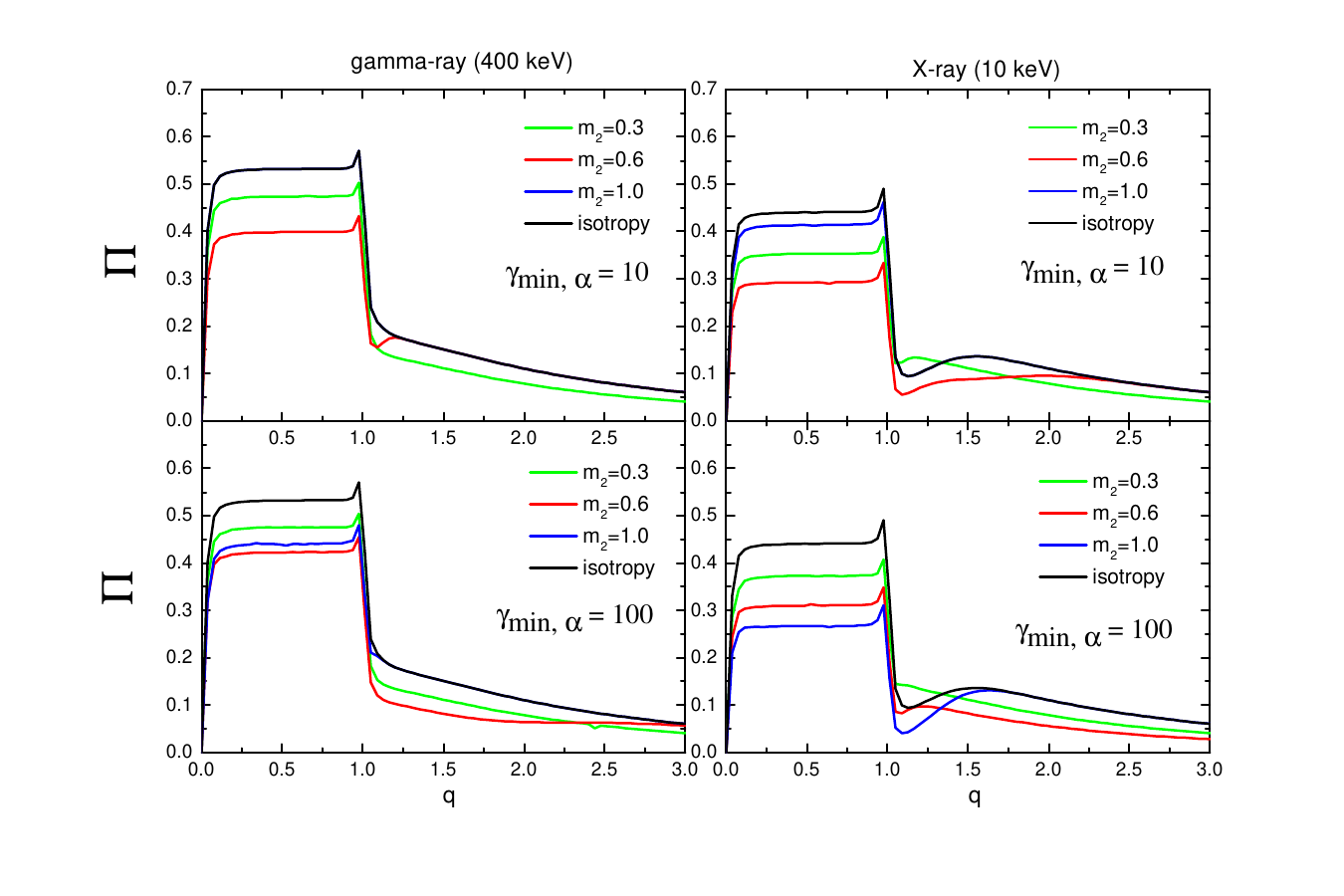}
\caption{The distribution of the PDs as a function of normalized viewing angles ($q$) for various $m_2$ in the $\gamma$-ray (the left panels) and X-ray (the bottom panels) bands. We take $\gamma_{min, \alpha}=10$ and $100$ in the upper and bottom panels, respectively. The black solid line represents the case of isotropic electron distribution. We fix the energy slope of $m_1=-0.3$ in this figure. \label{fig:res-q-pol}}
\end{figure}

To investigate the differences in the observed PD across different lines of sight, we calculated the PD in the $\gamma$-ray (400 keV) and X-ray bands (10 keV) as a function of the normalized viewing angle $q$ ($q=\theta_v/\theta_j$). Figure \ref{fig:res-q-pol} presents the time-averaged PD as a distribution over the normalized viewing angle for different values of $m_2$. For comparison, the PD produced by an isotropic electron distribution is also included. We can find that the PDs for the on-beaming ($q\lesssim 1$) case are much higher than those for the off-beaming ($q>1$) case in the two wavebands for both anisotropic and isotropic electron distributions. Additionally, for on-beaming ($q\lesssim 1$) case, the PD in the $\gamma$-ray band is $\sim 0.1$ higher than that in the X-ray band, regardless of whether the electrons are anisotropically or isotropically distributed. For all values of $m_2$ in this Figure, the PD in the $\gamma$-ray and X-ray bands produced by anisotropically distributed electrons is lower than that in the isotropic distribution case. Only when $m_2$ is large ($m_2 = 1$) does the PD under anisotropic conditions approach that of the isotropic scenario, since at large $m_2$ the radiation in these two bands is dominated by isotropically distributed electrons. This is consistent with our previous results. Moreover, when $m_2 = 1.0$, the PD at $\gamma_{min, \alpha}=10$ is approximately equal to that in the isotropic case, and it is higher than that at $\gamma_{min, \alpha}=100$. This is actually consistent with the results we obtained in Section \ref{subsec:pol-es}. It is because that when $\gamma_{min, \alpha}$ is smaller, the value of $\gamma_{iso}$ is smaller (see equation \ref{g_iso}). This will result in the radiation in the $\gamma$-ray and X-ray bands being dominated by isotropically distributed electrons. Therefore, the PD obtained under $\gamma_{min, \alpha}=10$ and $m_2=1$ is approximately equal to that in the isotropic case.

\begin{figure}[ht!]
\plotone{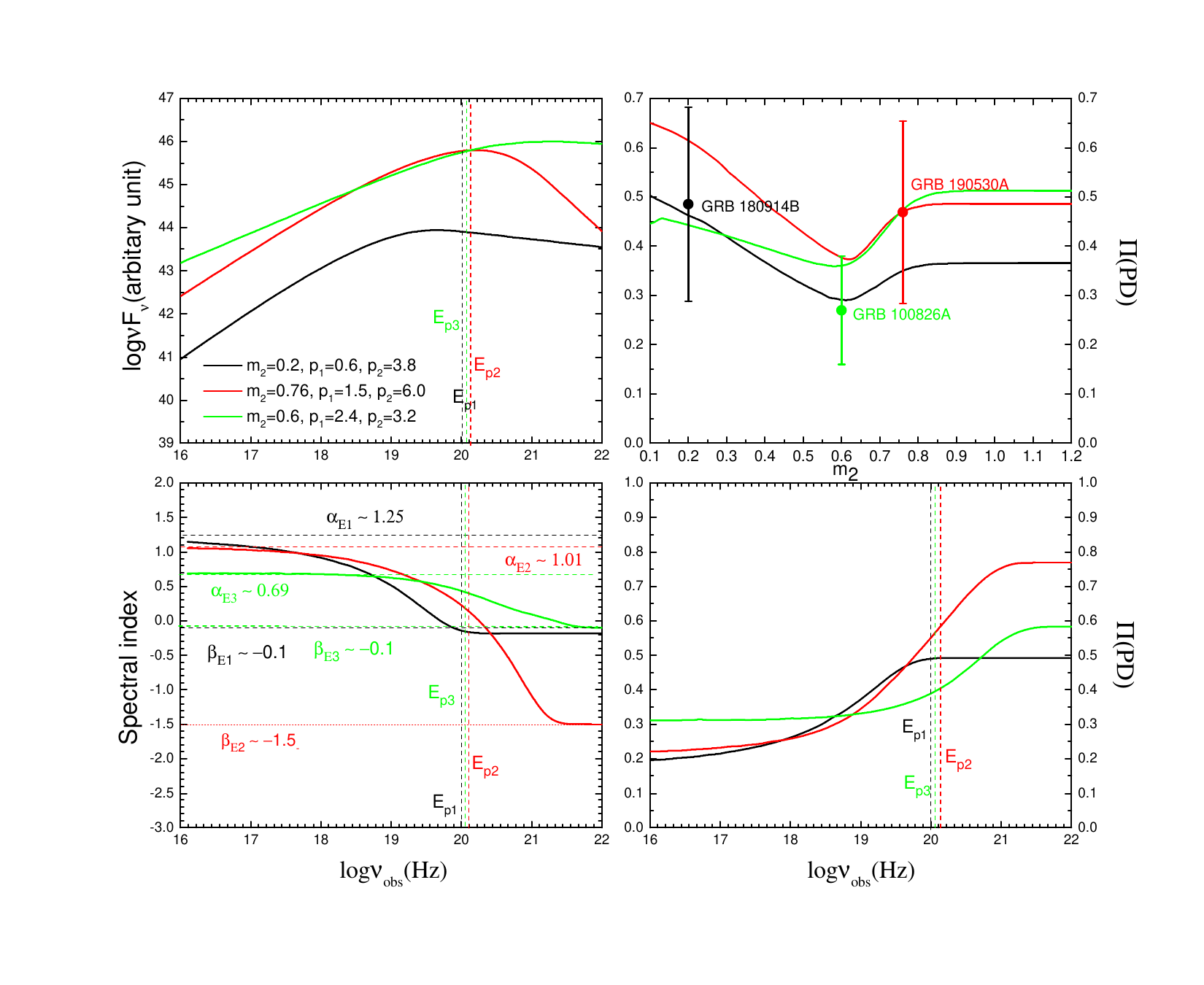}
\caption{The numerical results of the anisotropic electron distribution model compared to the PD and spectral data of GRBs 180914B, 190530A, and 100826A. The top-left panel shows the energy spectrum, and the bottom-left panel presents the corresponding spectral indices. The curves in the top-right panel display the PDs as a function of $m_2$, with data points representing the observed PDs of the bursts. The bottom-right panel shows the polarization spectra under the corresponding parameters. The black, red, and green curves correspond to GRBs 180914B, 190530A, and 100826A, respectively. The spectral and PD data of GRBs 180914B and 190530A are collected from \cite{Chattopadhyay+etal+2022}, and GRB 100826A is from \cite{2011ApJ...743L..30Y}. \label{fig:cp-data}}
\end{figure}

\subsection{Comparision with the measured PD and spectral data} \label{subsec:general}

To validate the anisotropic electron distribution model, we compare the numerical results derived from this model with the observed PD and spectral data. Our numerical results are compared with the measured PD ($\gamma$-ray band) and spectral data of GRBs 180914B, 190530A, and 100826A in Figure \ref{fig:cp-data}. GRBs 180914B and 190530A were detected by Cadmium Zinc Telluride Imager (CZTI) on board AstroSat with high PDs of $48.5_{-19.7}^{+19.7} \%$ and $46.9_{-18.5}^{+18.5}\%$, respectively \citep{Chattopadhyay+etal+2022}. GRB 100826A was detected by Gamma-Ray Burst Polarimeter (GAP) on board IKAROS with a PD of $27_{-11}^{+11} \%$ and was found to exhibit an evolution in its polarization angle \citep{2011ApJ...743L..30Y}. The measured PD and spectral data of these GRBs and their constrain to our model parameters are shown in Table \ref{tab: data}. Note that the spectral data consists of the low-energy spectral indices $\alpha_{E}$, the high-energy spectral indices $\beta_{E}$ ($\nu F_{\nu} \propto \nu^{\alpha_{E}}$ or $\propto \nu^{\beta_E}$), and the observed peak energy ($E_p$) of the energy spectrum. The comparison of the energy spectral indices is shown in the bottom left panel and the energy spectrum is shown in the top left panel. The curves in the top-right panel display the PDs as a function of $m_2$, with data points representing the observed PDs of the bursts. The energy-resolved PDs are shown in the bottom right panel. Considering that the energy slope of $m_1$ has almost no effect on the $\gamma$-ray band, we fix $m_1 = -0.3$ in the calculations of this figure. By adjusting the model parameters of $m_2$, $p_1$, and $p_2$, we obtained the results roughly consistent with the PD and spectral data. The detailed model parameters see Table \ref{tab: data}. GRBs 190530A and 100826A exhibit relatively high $m_2$ values ($0.76$ and $0.6$, respectively), whereas GRB 180914B shows a much lower value ($0.2$). This suggests that the electron distributions in GRBs 190530A and 100826A have a higher degree of anisotropy, while that in GRB 180914B is less anisotropic. In summary, our results demonstrate that, with the chosen model parameters, the theoretically derived PDs are roughly consistent with the observational data. Meanwhile, the spectral characteristics also show agreement with the observed. Hence, the anisotropic electron distribution model may provide a potential explanation for the polarization and spectral properties of some GRBs.

\section{Summary and discussion} \label{sec: conclusion}
In this paper, we calculate the polarization and radiation spectra of GRB prompt emission produced by anisotropically distributed electrons, and compared them with those in the case of isotropically distributed electrons. Our main results are summarized as follows.

\begin{enumerate}
\item The synchrotron PDs in the $\gamma$-ray and X-ray bands produced by anisotropically distributed electrons are systematically lower than those produced by isotropically distributed electrons, while the PD in the optical band could be either lower or higher than that of isotropically distributed electrons, depending primarily on the specific values of $m_1$, $m_2$, and $\gamma_{min, \alpha}$.

\item The energy slope of $m_1$ typically affects only the PD around the optical band, whereas $m_2$ influences the PDs in the gamma-ray, X-ray, and optical bands. Compared with isotropically distributed electrons, $m_1$ increases the PD around the optical band, while $m_2$ reduces the PDs in the gamma-ray, X-ray, and optical bands.

\item The radiation spectrum produced by anisotropically distributed electrons is harder than that of isotropic electrons. In particular, when $m_2\approx 1$, the spectral index in the low-energy band approaches zero, which is consistent with the Band spectrum. This suggests that the anisotropic electron distribution model may offer a potential explanation for the synchrotron fast cooling problem.

\item The anisotropically distributed electron model may provide a potential explanation for the PD and spectral data of some GRBs, such as GRB 180914B, 190530A, and 100826A.

\end{enumerate}

Synchrotron radiation is a leading candidate mechanism for GRB prompt emission. However, the observed low-energy spectral index ($\alpha_0 \sim 0$) deviates from the fast-cooling prediction of $1/2$ under the typical isotropic electron model. Our calculations indicate that the low-energy negative spectral index of isotropically distributed electrons is $\sim 1/2$ (see Figure \ref{fig:nu-chm2-pol}), which is consistent with the typical synchrotron spectrum in the fast cooling regime. While the radiation spectrum produced by anisotropically distributed electrons is harder than that of isotropic electrons. In particular, for $m_2 \approx 1$, the low-energy negative spectral index $\alpha_0$ of the radiation produced by anisotropically distributed electrons is $\sim 0$ (see Figure \ref{fig:nu-chm2-pol}), which is in agreement with the observed GRB spectra. Therefore, our results suggests that the anisotropic electron distribution model may offer a potential explanation for the synchrotron fast cooling problem.

Based on the numerical simulations, high-energy electrons ($\gamma>\gamma_{iso}$) follow an isotropic distribution, while lower-energy electrons ($\gamma < \gamma_{iso} $) follow an energy-dependent anisotropic distribution. Considering our results, the PD (above the X-ray band) produced by anisotropically distributed electrons is systematically lower than that produced by isotropically distributed electrons, and the difference in PD between the two models increases with the degree of anisotropy. Hence, the isotropically distributed electron model may explain the PD of high-energy emission (arising from electrons with $\gamma>\gamma_{iso}$), but struggles to account for the PD of low-energy emission (arising from electrons with $\gamma < \gamma_{iso}$). Therefore, the isotropically distributed electron model can not simultaneously account for the PD in both high- and low-energy bands. This suggests that the pitch-angle distribution of electrons can be constrained through polarization spectroscopy or multi-band polarization observations. This could potentially be tested by the future polarimeter POLAR-2 \citep{Kole+etal+2024, 2024ApJ...960...87F, 2026MNRAS.tmp..396G}, using its High-energy Polarization Detector (HPD) and Low-energy Polarization Detector (LPD). For the same GRB, if the isotropically distributed electron model fails to simultaneously explain the PD data measured by both HPD and LPD under the same set of parameters, then it may indicate that the pitch-angle distribution of electrons is more consistent with the anisotropic case. 

In addition to joint multi-band observations, combining polarization with radiation spectral features can help us impose more stringent constraints on the electron pitch-angle distribution. Hence, we also compared our numerical results with the observed polarization and spectral data of GRBs, and found that the anisotropic electron distribution model may simultaneously explain the PDs and spectral properties of some bursts, such as GRB 180914B, 190530A, and 100826A. Based on our results, the values of $m_2$ vary for different bursts, indicating differences in the degree of electron distribution anisotropy in different GRBs. These results need to be jointly tested in combination with the radiation spectrum and polarization of GRBs. The radiation spectrum can be provided by GRB detectors, such as Fermi-GBM, Svom, and Einstein Probe. However, polarization observations of GRBs are still relatively scarce, particularly high-precision polarization measurements. The launch of future polarization detectors will advance research in the field of GRB polarization, such as the POLAR-2 and eXTP \citep{2019SCPMA..6229502Z, 2025SCPMA..6819502Z} projects. After their launch, it will enable multi-wavelength joint tests of polarization models. This will help us impose stricter constraints on the pitch-angle distribution of electrons. 

Furthermore, this work focuses exclusively on linear polarization, following the standard synchrotron framework in which circular polarization (CP) is typically small for an isotropic pitch-angle distribution. However, an anisotropic pitch-angle distribution can generate CP at a higher level \citep{2023ApJ...959..137C}. The detectable CP is primarily expected in the low-energy bands \citep{1969SvA....13..396S}, such as the optical and radio bands, whereas current GRB polarimeters in the $\gamma$-ray and X-ray bands are only designed to measure linear polarization. We will perform the CP analysis with electron pitch-angle distribution for prompt GRB emission in the optical band in the future.  

We should note that our calculations adopt a uniform top-hat jet model. Realistic GRB jets may possess angular structures in both kinetic energy and LF \citep{2002ApJ...571..876Z, 2004MNRAS.354...86R}. In structured jets, the energy-dependent pitch-angle anisotropy would couple with the angularly varying energy and MF orientation across the jet structure. A quantitative exploration would require dedicated numerical calculations, but qualitatively, we expect that the relative suppression of PD due to pitch-angle anisotropy (compared to the isotropic case) might persist in structured jets, as the intrinsic emission from each fluid element is similarly affected by the local electron anisotropy.

Although this paper focuses on GRB prompt emission, the underlying microphysics of energy-dependent pitch-angle anisotropy might take effects on relativistic outflows in general. The Imaging X-ray Polarimetry Explorer (IXPE) has measured X-ray PDs of $\sim 10\%-20\%$ in several blazars \citep{2022ApJ...938L...7D, 2024A&A...689A.119K, 2024ApJ...963....5E}, values that are markedly lower than the $\sim 60\%-70\%$ maximum expected for a perfectly ordered MF and isotropic electrons. Our results demonstrate that pitch-angle anisotropy alone can systematically suppress the observed PDs. Therefore, the moderate PDs observed by IXPE may naturally arise from a combination of ordered fields and anisotropic electron populations. A dedicated study extending the current framework to blazar emission regions would be valuable.

\begin{acknowledgments}
This work is supported by the Guangxi Natural Science Foundation (2024GXNSFBA010350), the Scientific Research Project of Guangxi Minzu University (2021KJQD03), the College Students' Innovative Entrepreneurial Training Plan Program (S202410608023), the National Key R\&D Program of China (2023YFE0101200), the National Natural Science Foundation of China (Nos. 12473048, 12393811), the special funding for Guangxi Bagui Scholars, the Guangxi Science and Technology Innovation Platform Program (Leitai Action Plan, Grant No. Guike LT2600640026), Guangxi Key R\&D Program (Guangxi Funeng Action Plan, Grant No. Guike FN2504240040), and the "Guangxi Highland of Innovation Talents" Program. J.M is supported by the Yunnan Revitalization Talent Support Program (YunLing Scholar Project).
\end{acknowledgments}

\begin{deluxetable}{ccccccccc}[h] \label{tab: data}
	\tablecolumns{9}
	\setlength{\tabcolsep}{9pt}
	\tablewidth{0pc}
	\tablecaption{The measured PD ($\gamma$-ray band ) and spectral data of GRBs and their constrain to our model parameters.}
	\tabletypesize{\scriptsize}
	\tablehead{
		\colhead{GRB}&
		\colhead{PD(\%)}&
		\colhead{$\alpha_{E}/\beta_{E}$}&
		\colhead{$E_{\rm{p}}$ (keV)}&
		\colhead{Instrument (PD)}&
		\colhead{$m_2$} &
		\colhead{$p_1$}&
		\colhead{$p_2$} 
	}
	\startdata
	180914B & $48.5_{-19.7}^{+19.7}$ &$1.25_{-0.04}^{+0.04} / -0.1_{-0.70}^{+0.08}$ &$453_{-24}^{+26}$  & AstroSat-CZTI & $ 0.2$&$0.6$ &$3.8$  \\
	190530A & $46.9_{-18.5}^{+18.5}$ &$1.01_{-0.002}^{+0.022} / -1.50_{-0.25}^{+0.25}$ &$888_{-8}^{+8}$  & AstroSat-CZTI & $ 0.76$ & $1.5$ &$2.4$ \\
	100826A& $27_{-11}^{+11}$ &$0.69_{-0.05}^{+0.06} / -0.1_{-0.2}^{+0.1}$ &$606_{-109}^{+134}$ & IKAROS-GAP & $0.6$ & $2.4$ &$ 3.2$ \\
		\enddata
	\tablecomments{The spectral and PD data of AstroSat-CZTI GRBs and IKAROS-GAP GRB are collected from \cite{Chattopadhyay+etal+2022} and \cite{2011ApJ...743L..30Y}, respectively. The model parameters ($m_2$, $p_1$, and $p_2$) are given by comparing the PD data with the numerical results of our model.}
\end{deluxetable}

\bibliography{ms}{}
\bibliographystyle{aasjournal}



\end{document}